\documentclass[aps,prd,twocolumn,superscriptaddress,floatfix,nofootinbib]{revtex4}
\usepackage{longtable}
\usepackage{bm}
\usepackage{color}
\usepackage{graphicx}
\newcommand{\be}{\begin{equation}}
\newcommand{\ee}{\end{equation}}
\newcommand{\een}{\end{subequations}}
\newcommand{\ben}{\begin{subequations}}
\newcommand{\beq}{\begin{eqalignno}}
\newcommand{\eeq}{\end{eqalignno}}

\newcommand{\lsim}{\mathrel{\mathop{\kern 0pt \rlap
      {\raise.2ex\hbox{$<$}}}\lower.9ex\hbox{\kern-.190em $ \sim$}}}
\newcommand{\gsim}{\mathrel{\mathop{\kern 0pt
      \rlap{\raise.2ex\hbox{$>$}}}\lower.9ex\hbox{\kern-.190em $\sim$}}}

\newcommand {\vmin}{\mbox{$v_{min}$}}

\newcommand{\GeV}{{\rm GeV}}
\newcommand{\vesc}{v_{\rm esc}}
\newcommand{\Nesc}{N_{\rm esc}}
\newcommand{\vobs}{v_E}
\newcommand{\erf}{\mbox{erf}}
\newcommand{\Ed}{E^{\prime}}
\newcommand{\ud}{\text{d}}

\newcommand{\eR}{\mathcal{R}}

\begin{document}	

\title{Proton--philic spin--dependent inelastic Dark Matter (pSIDM) as
  a viable explanation of DAMA/LIBRA--phase2}
\author{Sunghyun Kang}
\email{francis735@naver.com}
\affiliation{Department of Physics, Sogang University, 
Seoul, Korea, 121-742}

\author{S. Scopel}
\email{scopel@sogang.ac.kr}
\affiliation{Department of Physics, Sogang University, 
Seoul, Korea, 121-742}
\author{Gaurav Tomar}
\email{tomar@sogang.ac.kr}
\affiliation{Department of Physics, Sogang University, 
Seoul, Korea, 121-742}
\author{Jong-Hyun Yoon}
\email{jyoon@sogang.ac.kr}
\affiliation{Department of Physics, Sogang University, 
Seoul, Korea, 121-742}

\date{\today}

\begin{abstract}
We show that the Weakly Interacting Massive Particle scenario of
proton-philic spin-dependent inelastic Dark Matter (pSIDM) can still
provide a viable explanation of the observed DAMA modulation amplitude
in compliance with the constraints from other experiments after the
release of the DAMA/LIBRA--phase2 data and including the recent bound
from COSINE--100, that uses the same $NaI$ target of DAMA. The pSIDM
scenario provided a viable explanation of DAMA/LIBRA--phase1 both for
a Maxwellian WIMP velocity distribution and in a halo--independent
approach. At variance with DAMA/LIBRA--phase1, for which the
modulation amplitudes showed an isolated maximum at low energy, the
DAMA/LIBRA--phase2 spectrum is compatible to a monotonically
decreasing one. Moreover, due to its lower threshold, it is sensitive
to WIMP--iodine interactions at low WIMP masses. Due to the
combination of these two effects pSIDM can now explain the yearly
modulation observed by DAMA/LIBRA only when the WIMP velocity
distribution departs from a standard Maxwellian. In this case the WIMP
mass $m_{\chi}$ and mass splitting $\delta$ fall in the approximate
ranges 7 GeV $\lsim m_{\chi}\lsim$ 17 GeV and 18
keV$\lsim\delta\lsim$29 keV. The recent COSINE--100 bound is naturally
evaded in the pSDIM scenario due to its large expected modulation
fractions. 
\end{abstract}

\pacs{95.35.+d,95.30.Cq}

\maketitle


\maketitle

\section{Introduction}
\label{sec:introduction}

About one quarter of the total mass density of the
Universe~\cite{planck} and more than 90\% of the halo of our Galaxy
are believed to be constituted by Dark Matter (DM) and Weakly
Interacting Massive Particles (WIMPs) are one of the most popular
candidates to compose it. The scattering rate of DM WIMPs in a
terrestrial detector is expected to present a modulation with a period
of one year due to the Earth revolution around the
Sun~\cite{Drukier:1986tm}.

The DAMA collaboration~\cite{dama_2008,dama_2010,dama_2013} has been
measuring for more than 15 years a yearly modulation effect in their
sodium iodide target. Such effect has a statistical significance of
more than $9\sigma$ and is consistent with what is expected from DM
WIMPs. However, in the most popular WIMP scenarios the DAMA modulation
appears incompatible with the results from many other DM experiments
that have failed to observe any signal so far.

This has lead to extend the class of WIMP models. In particular, one
of the few phenomenological scenarios that have been shown to explain
the DAMA effect in agreement with the constraints from other
experiments is proton--philic spin--dependent inelastic Dark Matter
(pSIDM)~\cite{psidm_2015,psidm_2017} for WIMP masses 10 $\GeV\lsim
m_{\chi}\lsim$ 30 $\GeV$ and a mass splitting 10 keV $\lsim
\delta\lsim$ 30 keV.

Recently the DAMA collaboration has released first result from the
upgraded DAMA/LIBRA-phase2 experiment~\cite{dama_2018}. Compared to
the previous data the two most important improvements are that now the
exposure has almost doubled and that the energy threshold has been
lowered from 2 keV electron--equivalent (keVee) to 1 keVee.  Moreover,
an important difference with the result of DAMA/LIBRA--phase1 is that
the new DAMA/LIBRA--phase2 spectrum of modulation amplitudes no longer
shows a maximum, but is rather monotonically decreasing with
energy\footnote{With the new DAMA data the goodness of fit of a
  standard Spin--Independent interaction and a Maxwellian velocity
  distribution has considerably worsened compared to
  DAMA-phase1~\cite{dama_phase2_first_analysis,freese_2018}. This can
  be alleviated assuming non--relativistic effective
  interactions~\cite{dama_2018_sogang} or two--component DM
  models~\cite{two_comp_dm}.}.  Moreover, several direct WIMP searches
have improved their bounds, including the COSINE--100
collaboration~\cite{cosine_nature} that has recently published an
exclusion plot about a factor of two below the DAMA region using 106
kg of $NaI$, the same target of DAMA, assuming an elastic,
spin--independent isoscalar WIMP nucleus interaction and a WIMP
Maxwellian velocity distribution. In light of these differences in the
present paper we wish to update the assessment of pSIDM with the new
DAMA/LIBRA--phase2 data, both in a scenario where the WIMP speed
distribution $f(v)$ is given by a standard Maxwellian and using a
halo--independent approach where $f(v)$ is not fixed.

In the present paper we will show that pSIDM can still provide a
viable explanation of the modulation effect after DAMA/LIBRA-phase2.
In particular, while the pSIDM scenario was able to explain
DAMA/LIBRA--phase1 both for a Maxwellian $f(v)$ and in a
halo--independent approach~\cite{psidm_2015,psidm_2017} in the present
paper we will show that for a Maxwellian WIMP velocity distribution it
provides a poor fit to the new DAMA data and for a range of the pSIDM
parameters in tension with the null results of other DM searches.  On
the other hand in a halo--independent approach the pSIDM scenario is
still viable. Moreover, we will show that the recent COSINE--100 bound
is naturally evaded in the pSDIM scenario due to its large expected
modulation fractions.

The paper is organized as follows. In Section~\ref{sec:psidm} we
outline the main features of the pSIDM scenario; in Section~
\ref{sec:maxwellian} we analyze the DAMA data adopting a standard
Maxwellian for the WIMP velocity distribution; in
Section~\ref{sec:halo_indep} we analyze the DAMA data in a
halo--independent approach.
In appendix\ref{sec:conclusions} we provide some details on how the
experimental constraints on pSIDM have been obtained.

\section{The pSIDM scenario}
\label{sec:psidm}
The most stringent bounds on an interpretation of the DAMA effect in
terms of WIMP--nuclei scatterings are obtained by detectors using
xenon (XENON1T~\cite{xenon_2018}, PANDAX--II~\cite{panda_2017},
LUX~\cite{lux_complete}) and germanium
(CDMS~\cite{cdms_ge,cdms_lite,super_cdms,cdms_2015}) whose spin is
mostly originated by an unpaired neutron while, on the other hand,
both sodium and iodine in DAMA have an unpaired proton.  This implies
that if the WIMP particle interacts with ordinary matter predominantly
via a spin--dependent coupling which is suppressed for neutrons it can
explain the DAMA effect in compliance with xenon and germanium
bounds~\cite{spin_n_suppression,spin_gelmini}. Actually, present
limits from xenon detectors require to tune the neutron/proton
coupling ratio $c^n/c^p$ to a small but non--vanishing
value~\cite{psidm_2015}. In the following we will adopt the
xenon--phobic combination $c^n/c^p$=-0.028, that minimizes the xenon
spin--dependent response using the nuclear structure functions
in~\cite{haxton2}~\footnote{The value $c^n/c^p$=-0.028 minimizes the
  XENON1T rate in the ROE 3 PE$<S_1<$ 70 PE~\cite{xenon_2018} for the
  Maxwellian case (using the nuclear form factors of
  Ref.~\cite{haxton2}). In the spin-dependent case the momentum
  dependence of the form factors is mild, so that this value is
  optimal in all the parameter space relevant to the pSIDM scenario.}.
This scenario is still constrained by droplet detectors and bubble
chambers (COUPP~\cite{coupp}, PICASSO~\cite{picasso},
PICO-60~\cite{pico60})) which all use nuclear targets with an unpaired
proton ($^{19}F$ and/or $^{127}I$).  As a consequence, this class of
experiments rules out a DAMA explanation in terms of WIMPs with a
spin--dependent coupling to
protons~\cite{spin_gelmini,pico2l,psidm_2015}.

In Ref.~\cite{psidm_2015} Inelastic Dark Matter~\cite{inelastic} (IDM)
was proposed to reconcile the above scenario to fluorine detectors. In
IDM a DM particle $\chi_1$ of mass $m_{\chi_1}=m_{\chi}$ interacts
with atomic nuclei exclusively by up--scattering to a second heavier
state $\chi_2$ with mass $m_{\chi_2}=m_{\chi}+\delta$. A peculiar
feature of IDM is that there is a minimal WIMP incoming speed in the
lab frame matching the kinematic threshold for inelastic upscatters
and given by:

\begin{equation}
v_{min}^{*}=\sqrt{\frac{2\delta}{\mu_{\chi N}}},
\label{eq:vstar}
\end{equation}

\noindent with $\mu_{\chi N}$ the WIMP--nucleus reduced
mass. This quantity corresponds to the lower bound of the minimal
velocity $v_{min}$ (also defined in the lab frame) required to deposit
a given recoil energy $E_R$ in the detector:

\begin{equation}
v_{min}=\frac{1}{\sqrt{2 m_N E_R}}\left | \frac{m_NE_R}{\mu_{\chi N}}+\delta \right |,
\label{eq:vmin}
\end{equation}

\noindent with $m_N$ the nuclear mass. In particular, indicating with
$v_{min}^{*Na}$ and $v_{min}^{*F}$ the values of $v_{min}^*$ for
sodium and fluorine, and with $v_{esc}$ the WIMP escape velocity,
constraints from WIMP--fluorine scattering events in droplet detectors
and bubble chambers can be evaded when the WIMP mass $m_{\chi}$ and
the mass gap $\delta$ are chosen in such a way that the hierarchy:

\begin{equation}
v_{min}^{*Na}<v_{esc}^{lab}<v_{min}^{*F},
\label{eq:hierarchy}
\end{equation}

\noindent is achieved, since in such case WIMP scatterings off
fluorine turn kinematically forbidden while those off sodium can
still serve as an explanation to the DAMA effect. So the pSIDM
mechanism rests on the trivial observation that the velocity
$v_{min}^*$ for fluorine is larger than that for sodium.

\section{Analysis}
\label{sec:analysis}

The expected rate in a given visible energy bin $E_1^{\prime}\le
E^{\prime}\le E_2^{\prime}$ of a direct detection experiment is given
by:

\begin{eqnarray}
R_{[E_1^{\prime},E_2^{\prime}]}(t)&=&MT_{exp}\int_{E_1^{\prime}}^{E_2^{\prime}}\frac{dR}{d
  E^{\prime}}(t)\, dE^{\prime} \nonumber\\
 \frac{dR}{d E^{\prime}}(t)&=&\sum_T \int_0^{\infty} \frac{dR_{\chi T}(t)}{dE_{ee}}{\cal
   G}_T(E^{\prime},E_{ee})\epsilon(E^{\prime})\label{eq:start2}\,d E_{ee} \nonumber\\
E_{ee}&=&q(E_R) E_R \label{eq:start},
\end{eqnarray}

\noindent with $\epsilon(E^{\prime})\le 1$ the experimental
efficiency/acceptance. In the equations above $E_R$ is the recoil
energy deposited in the scattering process (indicated in keVnr), while
$E_{ee}$ (indicated in keVee) is the fraction of $E_R$ that goes into
the experimentally detected process (ionization, scintillation, heat)
and $q(E_R)$ is the quenching factor, ${\cal
  G_T}(E^{\prime},E_{ee}=q(E_R)E_R)$ is the probability that the
visible energy $E^{\prime}$ is detected when a WIMP has scattered off
an isotope $T$ in the detector target with recoil energy $E_R$, $M$ is
the fiducial mass of the detector and $T_{exp}$ the live--time
exposure of the data taking.

In Eq.(\ref{eq:start}) the differential recoil rate $dR_{\chi T}(t)/dE_R$ is given by:

\be
\frac{d R_{\chi T}}{d E_R}(t)=\sum_T N_T\frac{\rho_{\mbox{\tiny WIMP}}}{m_{\mbox{\tiny WIMP}}}\int_{v_{min}}d^3 v_T f(\vec{v}_T,t) v_T \frac{d\sigma_T}{d E_R},
\label{eq:dr_de}
\ee

\noindent where $\rho_{\mbox{\tiny WIMP}}$ is the local WIMP mass
density in the neighborhood of the Sun (in the following we will
assume the standard value $\rho_{\mbox{\tiny WIMP}}$=0.3 GeV/cm$^3$),
$f(\vec{v}_T,t)$ is the WIMP velocity distribution (whose boost in the
Earth rest frame induces a time--dependence), $N_T$ the number of the
nuclear targets of species $T$ in the detector (the sum over $T$
applies in the case of more than one target), while:

\be \frac{d\sigma_T}{d E_R}=\frac{\sigma_0}{E_R^{max}}=\frac{2
  m_T}{4\pi v_T^2}\left [ \frac{1}{2 j_{\chi}+1} \frac{1}{2
    j_{T}+1}|\mathcal{M}_T|^2 \right ],
\label{eq:dsigma_de}
\ee

\noindent with $m_T$ the mass of the nuclear target, $j_{\chi}$=1/2
the spin of the WIMP, $E_R^{max}=2 \mu_{\chi T}^2/m_T v_T^2$ and
$\sigma_0$ the point--like WIMP--nucleon cross section. In the
following, for the calculation of the squared amplitude
$|\mathcal{M}_T|^2$ we will use the spin--dependent nuclear form
factors from~\cite{haxton2}\footnote{i.e., for the NREFT operator
  ${\cal O}_4$ in the notation of~\cite{haxton2}.} for all nuclei with
the exception of caesium and tungsten, for which we follow the same
procedure adopted in Appendix C of~\cite{sensitivities_2018}.

In particular, in each visible energy bin DAMA is sensitive to the
yearly modulation amplitude $S_m$, defined as the cosine transform of
$R_{[E_1^{\prime},E_2^{\prime}]}(t)$:

\begin{equation}
S_{m,[E_1^{\prime},E_2^{\prime}]}\equiv \frac{2}{T_0}\int_0^{T_0}
\cos\left[\frac{2\pi}{T_0}(t-t_0)\right]R_{[E_1^{\prime},E_2^{\prime}]}(t)dt,
\label{eq:sm}
\end{equation}  

\noindent with $T_0$=1 year and $t_0$=2$^{nd}$ June, while other
experiments put upper bounds on the time average $S_0$:

\begin{equation}
S_{0,[E_1^{\prime},E_2^{\prime}]}\equiv \frac{1}{T_0}\int_0^{T_0}
R_{[E_1^{\prime},E_2^{\prime}]}(t)dt.
\label{eq:s0}
\end{equation}

\subsection{Maxwellian analysis}
\label{sec:maxwellian}
In this Section we assume that the WIMP velocity distribution in the
Galactic rest frame is a standard isotropic Maxwellian at rest,
truncated at the escape velocity $\vesc$,
\begin{equation}
f_{\rm gal}(u) = \frac{1}{\pi^{3/2} v_0^3N_{\rm esc}} \, e^{-u^2/v_0^2} \, \Theta(\vesc - u ).
\end{equation}
Here $u$ is the WIMP speed in the Galactic rest frame, $v_0$ the
galactic rotational velocity at the Earth's position, $\Theta$ is the
Heaviside step function, and
\begin{equation}
\Nesc = \erf(z)-2 \, z \, e^{-z^2} / \pi^{1/2}
\end{equation}
with $z=\vesc/v_0$.  The WIMP speed distribution in the laboratory
frame can be obtained with a change of reference frame. It depends on
the speed of the Earth with respect to the Galactic rest frame, which
neglecting the ellipticity of the Earth orbit, is given by
\be
\vobs(t) = \big[ v_{\odot}^2 + v_{\oplus}^2 + 2 \, v_{\odot} \, v_{\oplus} \, \cos\gamma \, \cos[\omega(t-t_0)] \big]^{1/2}.
\label{eq:maxwellian}
\ee
In this formula, $v_{\odot}$ is the speed of the Sun in the Galactic
rest frame, $v_{\oplus}$ is the speed of the Earth relative to the
Sun, and $\gamma$ is the ecliptic latitude of the Sun's motion in the
Galaxy. We take $\cos\gamma \simeq 0.49$, $v_{\oplus}=2\pi(1~{\rm
  AU})/(1~{\rm year}) \simeq 29~{\rm km/s}$, $v_{\odot}=v_0+12~{\rm
  km/s}$, $v_0=220~{\rm km/s}$~\cite{v0_koposov}, and
$v_{esc}=550~{\rm km/s}$~\cite{vesc_2014}.

The velocity integral in Eq.~(\ref{eq:dr_de}) for the truncated Maxwellian
distribution is computed from the expression of the speed
distribution. We have obtained $S_0$ and $S_m$ by expanding it to
first order in $v_{\oplus}/v_{\odot}$.

To check how well pSIDM with a Maxwellian
distribution fits the DAMA/LIBRA--phase2 data $S^{\rm exp}_{{\rm m},k}
\pm \sigma_k$ in~\cite{dama_2018}, we perform a $\chi^2$ analysis
constructing the quantity
\begin{equation}
\chi^2(m_{\chi},\delta,\sigma_0)=\sum_{k=1}^{14} \frac{\left [S_{{\rm m},k}(m_{\chi},\delta,\sigma_0)-S^{\rm exp}_{{\rm m},k} \right ]^2}{\sigma_k^2},
  \label{eq:chi2}
  \end{equation}
where we consider 14 energy bins of width 0.5 keVee from 1 keVee to 8
keVee. 

\begin{figure}
\begin{center}
  \includegraphics[width=0.5\textwidth]{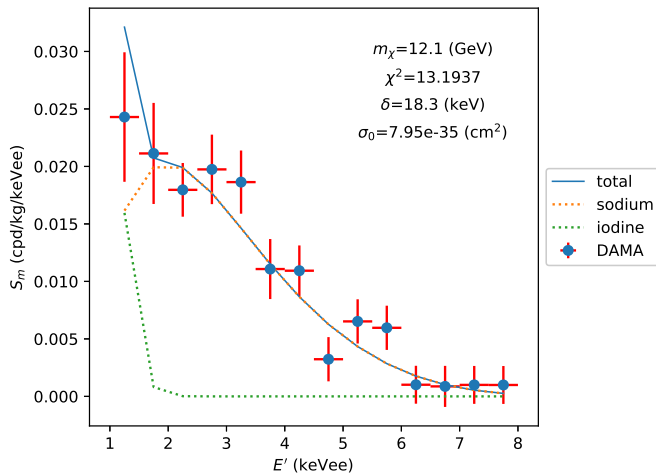}
  \caption{DAMA modulation amplitudes as a function of the measured
    ionization energy $\Ed$ for the absolute minimum of the pSIDM
    model in the case of a Maxwellian WIMP velocity distribution. The
    points with error bars correspond to the combined data of
    DAMA/NaI~\cite{dama_1998},
    DAMA/LIBRA--phase1~\cite{dama_2008,dama_2010} and
    DAMA/LIBRA--phase2~\cite{dama_2018}.}
\label{fig:sm_psidm}
  \end{center}
\end{figure}

The global minimum of $\chi^2(m_{\chi},\delta,\sigma_0)$ for pSIDM
occurs at $m_{\chi}=12.1~\GeV$, $\delta=18.3~{\rm keV}$, $\sigma_0 =
7.95 \times 10^{-35}~{\rm cm}^2$, and its value is
$\chi^2_{min}=13.19$ ($\text{$p$-value}=0.28$ with $14-3$ degrees of
freedom, which is an indication of a good fit). The modulation
amplitudes predicted by the pSIDM scenario are compared to the
combined data of DAMA/NaI~\cite{dama_1998},
DAMA/LIBRA--phase1~\cite{dama_2008,dama_2010} and
DAMA/LIBRA--phase2~\cite{dama_2018} in Fig.~\ref{fig:sm_psidm}.

The 5--$\sigma$ best-fit DAMA region in the ($m_{\chi}$--$\sigma_0$)
plane for the pSIDM scenario is compared to the corresponding 90\%
C.L. upper bounds from other DM searches in
Fig.~\ref{fig:psidm_mchi_sigma_exclusion} (see Appendix~\ref{app:exp}
for some details on how such constraints have been obtained). In the
same plot the IDM mass splitting is fixed to the absolute minimum of
the $\chi^2$, $\delta$=18.3 keV.  As can be seen from such figure the
DAMA effect is in strong tension with the upper bounds from PICO60,
KIMS and PICASSO.  On the other hand COSINE-100~\cite{cosine_nature},
that using the same $NaI$ target of DAMA has recently published an
exclusion plot about a factor of two below the DAMA modulation region
in the case of an elastic, spin--independent (SI) isoscalar
WIMP--nucleon interaction, does not exclude the pSIDM scenario. This
is due to the modulation fractions that in the pSIDM model are higher
than in the elastic case even for the case of a Maxwellian.  In fact
inelastic scattering is sensitive to the high--speed tail of the
velocity distribution due to the condition $v_{min}\ge v_{min}^*$. In
particular, the modulation residual measured by DAMA are at the level
of $S_m^{DAMA}\simeq$ 0.02 events/kg/day/keVee~\cite{dama_2018}, while
we estimate a bound from COSINE--100 $S_0^{COSINE}\lsim$0.13
events/kg/day/keVee after background subtraction (see
Appendix~\ref{app:exp}) on the non--modulated component of the
signal. This implies $S_m^{DAMA}/S_0^{DAMA}$ =
$S_m^{DAMA}/S_0^{COSINE}\times S_0^{COSINE}/S_0^{DAMA}$ $\gsim$ 0.12,
including a factor $S_0^{COSINE}/S_0^{DAMA}\simeq$ 0.8 due to a
difference between the energy resolutions and efficiencies in the two
experiments. For a standard Maxwellian WIMP velocity distribution in
the SI elastic case the predicted modulation fractions
$S_m^{DAMA}/S_0^{DAMA}$ are below such bound (for instance, for
$m_{\chi}$=10 GeV we find values of $S_m^{DAMA}/S_0^{DAMA}$ that range
between $\simeq$0.05 and $\simeq$0.12 for $E_{ee}<$ 3 keVee). In the
pSIDM case, however, the modulation fractions are all above such
value. For instance, for $m_{\chi}$=10 GeV and $\delta$=18 keV all the
modulation fractions for $E_{ee}<$ 6 keVee (i.e. in the range of the
DAMA signal) are above 0.4. This also explains why COSINE--100 does
not constrain the pSIDM scenario in the halo--independent approach of
the next Section.

We have also performed a combined fit including the upper bounds from
such experiments with the addition of COUPP and XENON1T and
COSINE--100, finding $\chi_{min}^2$=41.1 with a $p$--value
1.5$\times10^{-3}$ and 18 dof. Including $v_0$ and $u_{esc}$ as
nuisance parameters in the $\chi^2$ (we assume $v_0$=(220$\pm$ 20)
km/s~\cite{v0_koposov} and $u_{esc}$=(550$\pm$ 30)
km/s~\cite{vesc_2014}) does not improve the fit (we find
$\chi_{min}^2$=41.0). This confirms that, at variance with the
analyses of Ref.~\cite{psidm_2015,psidm_2017}, after the release of
the DAMA/LIBRA--phase2 data the pSIDM scenario in the Maxwellian case
is ruled out. There are two reasons for this.  The first reason is
that while the DAMA/LIBRA--phase1 data where only sensitive to
scattering events off sodium, the DAMA/LIBRA--phase2 data have a lower
threshold and are now also sensitive to scattering events off iodine
for $\Ed<$2 keVee at low WIMP masses. This makes it more difficult to
fit the model to the data since in the pSIDM scenario the scaling
between the cross sections off iodine and sodium is fixed (the
parameter $c^n/c^p$, that would allow to change such scaling is locked
to the combination that suppresses the response on xenon). Moreover,
in the scenario described in Section~\ref{sec:psidm} a minimal value
of the mass splitting parameter $\delta$ is required in order to
comply with the condition of Eq.(\ref{eq:hierarchy}), which, at the
same time automatically implies that the recoil energy
$E^*_R\equiv E_R(v_{min}^{*Na})$, and so a single maximum of the
modulation amplitude spectrum, falls inside the range of the DAMA
signal~\cite{psidm_2015} (the energy $E^*_R$ maximizes the velocity
integral in Eq.~(\ref{eq:eta_h})). Indeed, the DAMA/LIBRA--phase1 data
showed a single maximum in the 2.5 keVee$<\Ed<$3 keVee energy bin in
the measured modulation amplitudes~\cite{dama_2008,dama_2010},
implying an acceptable fit for the pSIDM model.  On the other hand the
DAMA/LIBRA--phase2 data show an energy spectrum of the modulation
amplitudes more compatible to a monotonically decreasing one, closer
to what expected for elastic scattering. As a consequence of this the
DAMA/LIBRA--phase2 $\chi^2$ pulls to low values of the $\delta$ mass
splitting (indeed, the Maxwellian best--fit configuration
$m_{\chi}=12.1~\GeV$, $\delta$=18.3 keV falls below the
halo--independent compatibility region discussed in the next Section
and shown in Fig.~\ref{fig:psidm_mchi_delta_halo_indep}), entering in
conflict with the requirement of Eq.(\ref{eq:hierarchy})\footnote{On
  the other hand, a fit of the DAMA/LIBRA--phase1 data below 8 keV to
  the pSIDM scenario yields $\chi^2_{min}$=8.6 with 12-3 dof,
  $m_{\chi}$=12.8 GeV, $\sigma_0$=4.5$\times$ 10$^{-34}$ cm$^2$ and
  $\delta$=23.6 keV, in agreement with the requirement of
  Eq.(\ref{eq:hierarchy}).}.
\begin{figure}
\begin{center}
\includegraphics[width=0.5\textwidth]{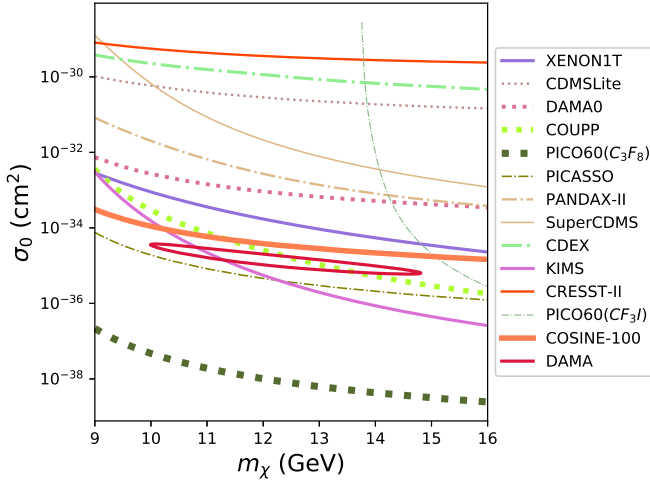}     
\end{center}
\caption{The 5--$\sigma$ best-fit DAMA region for the pSIDM scenario
  is compared to the corresponding 90\% C.L. upper bounds from other
  DM searches for a Maxwellian WIMP velocity distribution. In the plot
  the IDM mass splitting is fixed to $\delta$=18.3 keV, which
  corresponds to the absolute minimum of the $\chi^2$.}
\label{fig:psidm_mchi_sigma_exclusion}
\end{figure}
\subsection{Halo--independent analysis}
\label{sec:halo_indep}

In the halo--independent approach~\cite{factorization} the expected
rate in a direct detection experiment is recast in the
form~\cite{del_nobile_generalized}:

\be
\label{eq:R3}
R_{[\Ed_1, \Ed_2]}(t) = \int_0^\infty \ud\vmin \, \tilde{\eta}(\vmin, t) \,  \eR_{[\Ed_1, \Ed_2]}(\vmin) \, ,
\ee

\noindent where the dependence on astrophysics is contained in the
halo function:

\be
\tilde{\eta}(\vmin, t) = \frac{\rho_{\chi}}{m_{\chi}} \, \sigma_{0} \, \eta(\vmin,t),
\label{eq:eta_tilde_h}
\ee

\noindent and the WIMP velocity distribution is contained in the velocity integral:

\begin{equation}
\eta(\vmin,t)=\int_{\vmin}^\infty \ud v \, \, \frac{f(v, t)}{v},
   \label{eq:eta_h}
\end{equation}

\noindent while the response function $\eR_{[\Ed_1, \Ed_2]}(\vmin)$ is given by:

\begin{eqnarray}
  &&\eR_{[\Ed_1, \Ed_2]}(\vmin)=\sum_T N_T \frac{v_T^2}{\sigma_0}\frac{d\sigma_T}{d E_R}\times\nonumber\\
  &&\int_{\Ed_1}^{\Ed_2}d\Ed \epsilon(\Ed)
{\cal G}_T(E^{\prime},E_{ee}(\vmin).
\label{eq:RT}
\end{eqnarray}

\noindent Notice that for a standard spin--dependent interaction the
scattering amplitude in Eq.(\ref{eq:dsigma_de}) does not depend on $v_T$
so the term $v_T^2$ in the equation above cancels out in the product
$v_T^2 d\sigma_T/d E_R$.
\begin{figure}
\begin{center}
\setbox1=\hbox{\includegraphics[height=5.8cm]{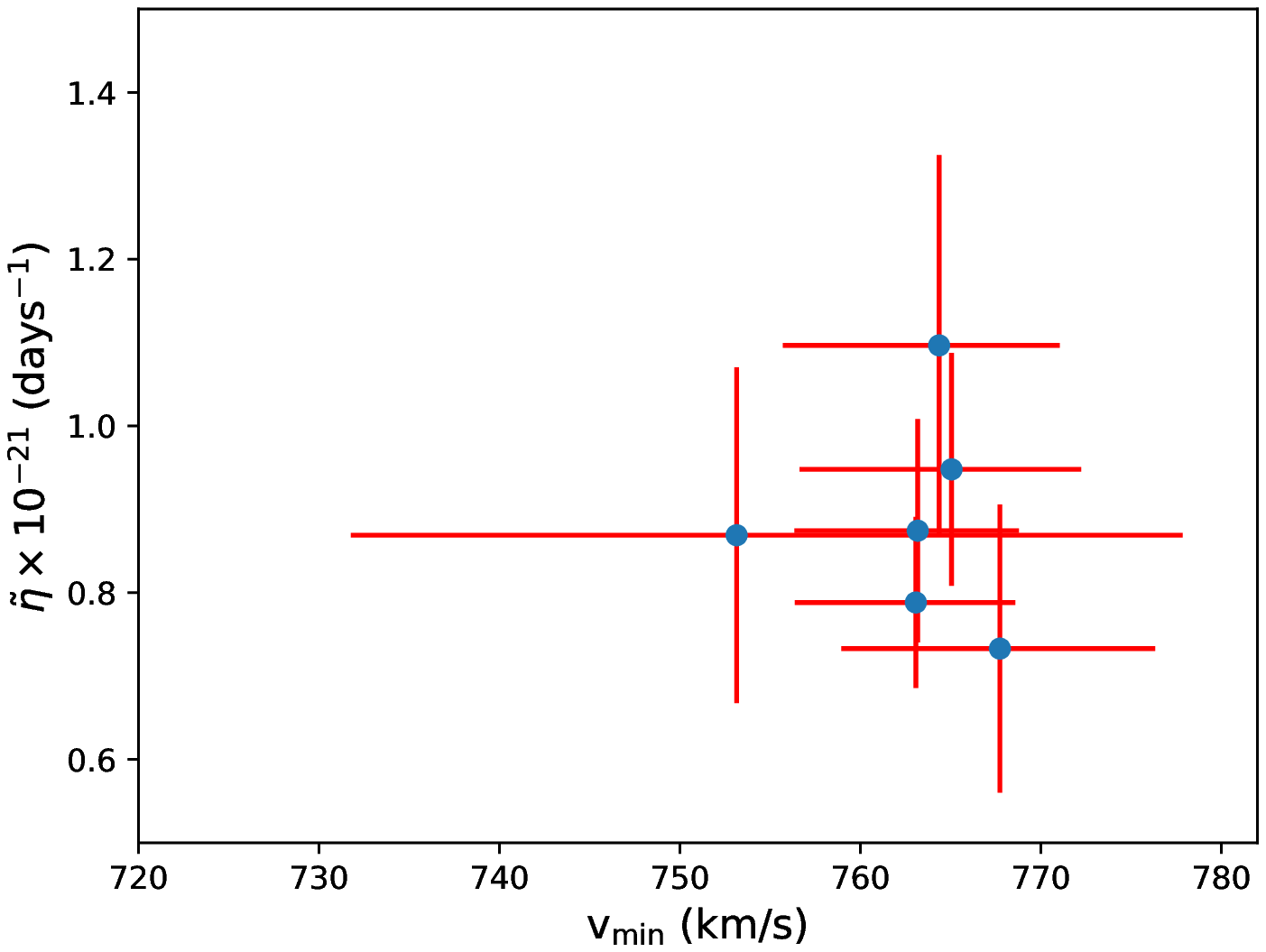}}
\includegraphics[height=6.7cm]{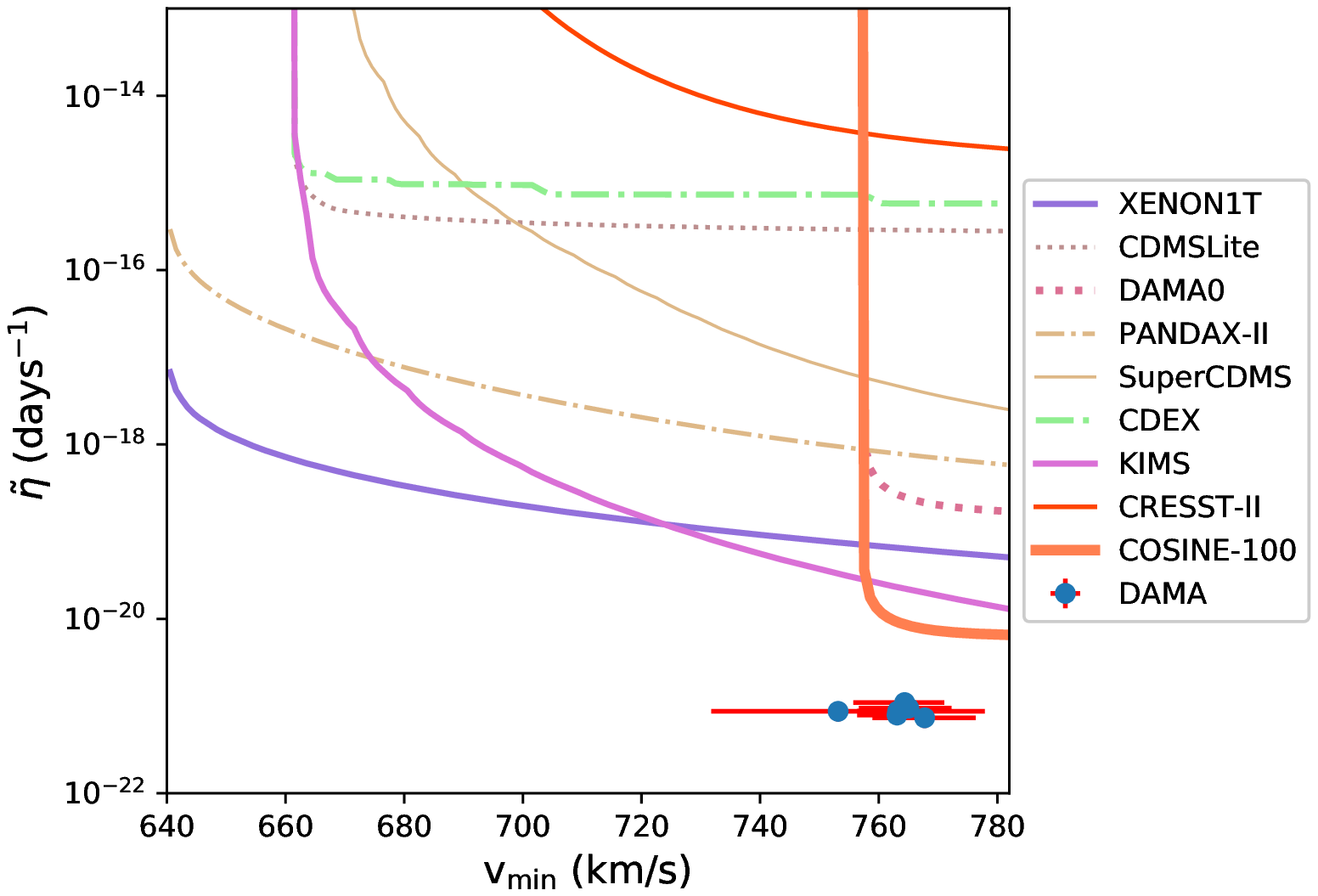}\llap{\makebox[\wd1][l]{\raisebox{0.9cm}{\includegraphics[height=2.cm,bb=75 223 626 626]{figs/fig3b}}}}
\end{center}
\caption{Measurements of $\overline{\tilde{\eta}^1}_{[v_{{\rm
          min},1},v_{{\rm min},2}]}$ (DAMA/LIBRA) and upper bounds
  $\tilde{\eta}^{\rm lim}$ (same experiments as in
  Fig.~\ref{fig:psidm_mchi_sigma_exclusion}) for pSIDM in the
  benchmark point $m_{\chi}$=11.4 GeV, $\delta$=23.7 keV.}
\label{fig:vmin_eta_exclusion_benchmark}
\end{figure}
Due to the revolution of the Earth around the Sun, the velocity
integral $\tilde{\eta}(\vmin, t)$ shows an annual modulation that can
be approximated by the first terms of a harmonic series,
\begin{equation}
\label{etat}
\tilde{\eta}(\vmin, t) = \tilde{\eta}^0(\vmin) +
\tilde{\eta}^1(\vmin) \, \cos\!\left[ \omega (t - t_0) \right],
\end{equation}

\noindent with the only requirement that
$|\tilde{\eta}^1|\le\tilde{\eta}^0$.  In this approach measured rates
$R^i_{[\Ed_1, \Ed_2]}$ (with $i=0,1$) are mapped into suitable
averages of the two halo functions $\tilde{\eta}^i$. Averages
$\overline{\tilde{\eta}^i}_{[v_{{\rm min},1}, v_{{\rm min},2}]}$
($i=0,1$) using $\eR(\vmin)$ in Eq.~(\ref{eq:RT}) as a weight function
can then be directly obtained from the experimental
data $R^i_{[\Ed_1, \Ed_2]}$ as~\cite{del_nobile_generalized}:

\begin{eqnarray}
\overline{\tilde{\eta}^i}_{[v_{{\rm min},1},v_{{\rm min},2}]} &=&
  \frac{\int_0^\infty \ud\vmin \, \tilde{\eta}^i(\vmin) \, \eR_{[\Ed_1, \Ed_2]}(\vmin)}{\int_0^\infty \ud\vmin \, \eR_{[\Ed_1, \Ed_2]}(\vmin)} =\nonumber\\
&=& \frac{R^i_{[\Ed_1, \Ed_2]}}{\int_0^\infty \ud\vmin \, \eR_{[\Ed_1, \Ed_2]}(\vmin)},
\label{eq:eta_average}
\end{eqnarray}

\noindent The result of such procedure is shown in
Fig.~\ref{fig:vmin_eta_exclusion_benchmark}, where the determinations
of $\overline{\tilde{\eta}^1}_{[v_{{\rm min},1},v_{{\rm min},2}]}$
from DAMA/LIBRA--phase2 data are shown with error bars for the
benchmark point $m_{\chi}$=11.4 GeV, $\delta$=23.7 keV. 

The velocity intervals $[v_{{\rm min},1}, v_{{\rm min},2}]$ are
defined as those velocity intervals where the weight function
$\eR_{[\Ed_1, \Ed_2]}(\vmin)$ is sizeably different from zero.  In
particular, to determine the $\vmin$ interval corresponding to each
detected energy interval $[\Ed_1, \Ed_2]$ in DAMA we choose to use
$68\%$ central quantile intervals of the response function, i.e. we
determine ${\vmin}_{,1}$ and ${\vmin}_{,2}$ such that the areas under
the function $ \eR_{[\Ed_1, \Ed_2]}(\vmin)$ to the left of
${\vmin}_{,1}$ and to the right of ${\vmin}_{,2}$ are each separately
$16\%$ of the total area under the function. This gives the horizontal
width of the crosses corresponding to the rate measurements in
Fig.~\ref{fig:vmin_eta_exclusion_benchmark}. On the other hand, the
horizontal placement of the vertical bar in the crosses corresponds to
the average of $\vmin$, i.e., $\vmin(\text{vertical
  bar})=\big[\int_0^\infty \ud\vmin \,\vmin \eR_{[\Ed_1,
      \Ed_2]}(\vmin)\big]/\big[\int_0^\infty \ud\vmin \,
  \eR_{[\Ed_1, \Ed_2]}(\vmin)\big]$. The extension of the vertical bar
shows the $1 \sigma$ interval around the central value of the measured
rate.

To compute upper bounds on $\tilde{\eta}^{\,0}$ from
upper limits $R_{[\Ed_1, \Ed_2]}^{\rm lim}$ on the unmodulated rates,
we follow the conservative procedure in
Ref.~\cite{factorization}. Since $\tilde{\eta}^{\,0}(\vmin)$ is by
definition a non-decreasing function, the lowest possible
$\tilde{\eta}^0(\vmin)$ function passing through a point
$(v_0,\tilde{\eta}^{\,0})$ in $\vmin$ space is the downward step
function $\tilde{\eta}^0 \, \theta(v_0 - \vmin)$. The maximum value of
$\tilde{\eta}^0$ allowed by a null experiment at a certain confidence
level, denoted by $\tilde{\eta}^{\rm lim}(v_0)$, is then determined by
the experimental limit on the rate $R_{[\Ed_1, \Ed_2]}^{\rm lim}$ as
\begin{equation}
\tilde{\eta}^{\rm lim}(v_0) = \frac{R^{\rm lim}_{[\Ed_1,
      \Ed_2]}}{\int_0^{v_0} \ud \vmin \, \eR_{[\Ed_1, \Ed_2]}(\vmin)}\ .
\label{eq:etalim}
\end{equation}
The corresponding upper limits at 90\% C.L. are shown as continuous
lines in Figs.~\ref{fig:vmin_eta_exclusion_benchmark} for the same
experiments shown in Fig.~\ref{fig:psidm_mchi_sigma_exclusion}.

\begin{figure}
\begin{center}
\includegraphics[width=0.5\textwidth]{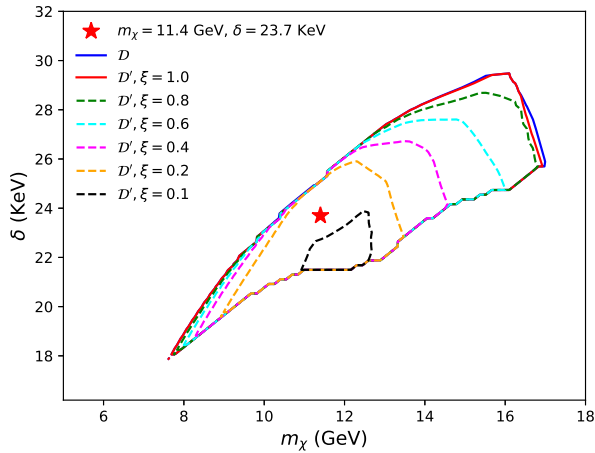}     
\end{center}
\caption{In the region inside the closed solid contour the
  compatibility parameter ${\cal D}$ defined in
  Eq.~(\ref{eq:compatibility_parameter}) is less than unity, implying
  that no 1--$\sigma$ interval of the quantities
  $\overline{\tilde{\eta}^1}_{[v_{{\rm min},1},v_{{\rm min},2}]}$
  obtained from the DAMA/LIBRA data lies completely above any of the
  upper bounds $\tilde{\eta}^{\rm lim}$. Inside the dashed closed
  contours the alternative compatibility factor ${\cal D}^{\prime}$
  defined in Eq.~(\ref{eq:compatibility_parameter_rates}) is less than
  unity for different values of the maximal modulation fraction $\xi$.
  The starred point corresponds to the benchmark shown in
  Fig.~\ref{fig:vmin_eta_exclusion_benchmark}.}
\label{fig:psidm_mchi_delta_halo_indep}
\end{figure}

For the specific benchmark $m_{\chi}$=11.4 GeV, $\delta$=23.7 keV
shown in Fig.~\ref{fig:vmin_eta_exclusion_benchmark} one can see that
pSIDM cannot be ruled out as an explanation of the DAMA/LIBRA effect
since in all the energy range of the signal one has
$|\overline{\tilde{\eta}^1}_{[v_{{\rm min},1},v_{{\rm
        min},2}]}|\ll\tilde{\eta}^{\rm lim}$. The same benchmark is
represented by a starred point in
Fig~\ref{fig:psidm_mchi_delta_halo_indep} and lies inside the the
closed contour where the compatibility factor defined
as~\cite{psidm_2015}:
\begin{equation}
  {\cal D}(m_{DM},\delta) \equiv \max_{i}
\left (\frac{\overline{\tilde{\eta}^1}_{[v^i_{{\rm min},1},v^i_{{\rm min},2}]}-\sigma_i}{\min_i \tilde{\eta}^{\rm lim}(v_0^i)} \right ),
\label{eq:compatibility_parameter}
\end{equation}

\noindent is less than unity. In the equation above $[v^i_{{\rm
      min},1},v^i_{{\rm min},2}]$ and $v_0^i$ represent intervals and
averages of $\vmin$ for each of the $i$=1...14 DAMA/LIBRA bins below
$\Ed$=8 keVee, while $\sigma_i$ is the 1--$\sigma$ fluctuation on
$\overline{\tilde{\eta}^1}_{[v^i_{{\rm min},1},v^i_{{\rm
        min},2}]}$. In particular, the requirement ${\cal
  D}(m_{DM},\delta)<$1 ensures that within the solid closed contour of
Fig.~\ref{fig:psidm_mchi_delta_halo_indep} no 1--$\sigma$ interval of
the quantities $\overline{\tilde{\eta}^1}_{[v_{{\rm min},1},v_{{\rm
        min},2}]}$ obtained from the DAMA/LIBRA data lies completely
above any of the upper bounds $\tilde{\eta}^{\rm lim}$.  In
Fig.~\ref{fig:psidm_mchi_delta_halo_indep} we also provide additional
dashed closed contours which correspond to an alternative, more
accurate definition of the compatibility factor: once the averages of
the modulated halo function $\overline{\tilde{\eta}^1}$ are determined
from the DAMA data, we determine a minimal set of averages of the
unmodulated halo function $\overline{\tilde{\eta}^0}$ complying with
the condition that $\overline{\tilde{\eta}^0}$ is a non--increasing
function of $v_{min}$ and that
$|\overline{\tilde{\eta}^1}|<\xi\overline{\tilde{\eta}^0}$ with
$\xi\le$1 ($\xi$=1 corresponding to 100\% modulation). We then use the
$\overline{\tilde{\eta}^0}$'s to calculate for different values of
$\xi$ the expected rate $R^{exp}_n(m_{DM},\delta,\xi)$ for each
experiment exp and energy bin $n$. Indicating with $R^{exp,lim}_n$ the
corresponding 90\% C.L. upper bound we adopt as compatibility factor:

\begin{equation}
  {\cal D}^{\prime}(m_{DM},\delta,\xi) \equiv \max_{exp,n}
\frac{R^{exp}_n(m_{DM},\delta,\xi)}{R^{exp,lim}_n},
\label{eq:compatibility_parameter_rates}
\end{equation}

\noindent where, again, ${\cal D}^{\prime}<$1 indicates compatibility.
As long as $\xi=1$, ${\cal D}$ and ${\cal D}^{\prime}$ yield similar
results (implying that, although the DAMA result requires all the
$\overline{\tilde{\eta}^1}$'s the bound is driven by only one of the
corresponding $\overline{\tilde{\eta}^0}$'s). In particular, from
Fig.~\ref{fig:psidm_mchi_delta_halo_indep} one can see that in a
halo--independent approach the pSIDM scenario can explain the
DAMA/LIBRA data for 7 GeV $\lsim m_{\chi}\lsim$ 17 GeV and 18
keV$\lsim\delta\lsim $29 keV.  As expected, when $\xi<1$ (smaller
modulation fractions) the compatibility region in
Fig.~\ref{fig:psidm_mchi_delta_halo_indep} shrinks. Inside the
smallest dashed contour of Fig.~\ref{fig:psidm_mchi_delta_halo_indep}
${\cal D}^{\prime}$ eventually reaches a plateau with minimum value
slightly lower than 0.1, implying the compatibility of modulation
fractions as low as $\simeq$0.1.


\section{Conclusions}
\label{sec:conclusions}

We have shown that the Weakly Interacting Massive Particle scenario of
proton-philic spin-dependent inelastic Dark Matter (pSIDM) can still
provide a viable explanation of the observed DAMA modulation amplitude
in compliance with the constraints from other experiments after the
release of the DAMA/LIBRA phase--2 data. The pSIDM scenario provided a
viable explanation of DAMA/LIBRA--phase 1 both for a Maxwellian WIMP
velocity distribution and in a halo--independent approach.  At
variance with DAMA/LIBRA--phase1, for which the modulation amplitudes
showed an isolated maximum at low energy, the DAMA/LIBRA--phase2
spectrum is compatible to a monotonically decreasing one. Moreover,
due to its lower threshold, it is sensitive to WIMP--iodine
interactions at low WIMP masses. Due to the combination of these two
effects pSIDM can now explain the modulation observed by DAMA/LIBRA
only when the WIMP velocity distribution departs from a standard
Maxwellian. In this case the WIMP mass $m_{\chi}$ and mass splitting
$\delta$ fall in the approximate ranges 7 GeV $\lsim m_{\chi}\lsim$ 17
GeV and 18 keV$\lsim\delta\lsim $29 keV. The recent COSINE--100 bound
is naturally evaded in the pSDIM scenario due to its large expected
modulation fractions, because inelastic scattering is sensitive to the
high--speed tail of the velocity distribution.

We conclude by pointing out that strictly speaking our analysis can
only lead to the conclusion that in the pSIDM scenario it is not
possible to rule out a DAMA explanation in terms of WIMPs in a
halo-independent way. On the other hand, the problem of inverting the
halo function $\tilde{\eta}^1$ to obtain a velocity distribution
$f(v)$ that, due to the boost from the Galactic to the lab rest
frames, leads to the correct time and energy dependence of the DAMA
signal is a complex and highly degenerate one that would require a
dedicated analysis beyond the scope of our paper. Moreover, due to the
very strong existing limits from null searches the pSIDM scenario
requires considerable tuning, such as a large suppression of the
spin--independent coupling, a specific range of the Galactic escape
velocity according to Eq. (3) and the tuning of the $c^n/c^p$
couplings ratio. As a consequence pSIDM appears challenging both from
the point of view of particle physics model--building and of that of
astrophysics. However, in spite of these challenges, at the very least
the pSIDM scenario can be considered as a proof of concept of the fact
that the parameter space of WIMP direct detection is wider than
usually assumed and that experimentally an explanation of the DAMA
signal in terms of WIMPs has not yet been completely probed.

\acknowledgments This research was supported by the Basic Science
Research Program through the National Research Foundation of Korea
(NRF) funded by the Ministry of Education, grant number
2016R1D1A1A09917964. 

\appendix
\section{Experiments}
\label{app:exp}

With the exception of the recent COSINE-100
result~\cite{cosine_nature} we fix the experimental input (exposure,
energy resolution, quenching factors, efficiency, measured count
rates, etc.)  for both the DAMA/LIBRA experiment and for other DM
searches as described in appendix B of~\cite{sensitivities_2018} and
appendix A of~\cite{anapole_2018}.  Recently~\cite{cosine_nature} the
COSINE--100 collaboration has published a bound about a factor of two
below the DAMA region using 106 kg of $NaI$, the same target of
DAMA. Such result assumes an elastic, spin--dependent isoscalar WIMP
nucleus interaction for a WIMP Maxwellian velocity distribution with
standard parameters, and relies on a Montecarlo~\cite{cosine_bck} to
subtract the different backgrounds of each of the eight crystals used
in the analysis. In Ref.~\cite{cosine_nature} the amount of residual
background after subtraction is not provided, and depends on the
expected signal shape, so should not in principle be used to constrain
a signal with a spectral shape different from the specific scenario
adopted in Ref.~\cite{cosine_nature}. However, especially in the
halo--independent case, for which the expected spectral shape cannot
be used to constrain the background, we deem that using the same
residual background of Ref.~\cite{cosine_nature} should lead to an
optimistic estimate of the bound.  So we have assumed for COSINE--100
a constant background $b$ at low energy (2 keVee$< E_{ee}<$ 8 keVee),
and we have estimated $b$ by tuning it to reproduce the exclusion plot
in Fig.4 of Ref.~\cite{cosine_nature} for the isoscalar
spin-independent elastic case. The result of our procedure yields
$b\simeq$0.13 events/kg/day/keVee, which implies a subtraction of
about 95\% of the background. We have then used the same value to
calculate the bounds for pSIDM. We take the energy resolution
$\sigma/\mbox{keV}=0.3171 \sqrt{E_{ee}/\mbox{keVee}}+0.008189
E_{ee}/\mbox{keVee}$ averaged over the COSINE--100
crystals~\cite{cosine_private} and the efficiency for nuclear recoils
from Fig.1 of Ref.~\cite{cosine_nature}. Quenching factors for sodium
and iodine are assumed to be equal to 0.3 and 0.09 respectively, the
same values used by DAMA.


\begin{thebibliography}{99}

\expandafter\ifx\csname natexlab\endcsname\relax\def\natexlab#1{#1}\fi
\expandafter\ifx\csname bibnamefont\endcsname\relax
  \def\bibnamefont#1{#1}\fi
\expandafter\ifx\csname bibfnamefont\endcsname\relax
  \def\bibfnamefont#1{#1}\fi
\expandafter\ifx\csname citenamefont\endcsname\relax
  \def\citenamefont#1{#1}\fi
\expandafter\ifx\csname url\endcsname\relax
  \def\url#1{\texttt{#1}}\fi
\expandafter\ifx\csname urlprefix\endcsname\relax\def\urlprefix{URL }\fi
\providecommand{\bibinfo}[2]{#2}
\providecommand{\eprint}[2][]{\url{#2}}

\bibitem[{\citenamefont{Ade et~al.}(2014)}]{planck}
\bibinfo{author}{\bibfnamefont{P.~A.~R.} \bibnamefont{Ade}}
  \bibnamefont{et~al.} (\bibinfo{collaboration}{Planck}),
  \bibinfo{journal}{Astron. Astrophys.} \textbf{\bibinfo{volume}{571}},
  \bibinfo{pages}{A16} (\bibinfo{year}{2014}), \eprint{1303.5076}.

\bibitem[{\citenamefont{Drukier et~al.}(1986)\citenamefont{Drukier, Freese, and
  Spergel}}]{Drukier:1986tm}
\bibinfo{author}{\bibfnamefont{A.~K.} \bibnamefont{Drukier}},
  \bibinfo{author}{\bibfnamefont{K.}~\bibnamefont{Freese}}, \bibnamefont{and}
  \bibinfo{author}{\bibfnamefont{D.~N.} \bibnamefont{Spergel}},
  \bibinfo{journal}{Phys. Rev.} \textbf{\bibinfo{volume}{D33}},
  \bibinfo{pages}{3495} (\bibinfo{year}{1986}).

\bibitem[{\citenamefont{Bernabei et~al.}(2008)}]{dama_2008}
\bibinfo{author}{\bibfnamefont{R.}~\bibnamefont{Bernabei}} \bibnamefont{et~al.}
  (\bibinfo{collaboration}{DAMA}), \bibinfo{journal}{Eur. Phys. J.}
  \textbf{\bibinfo{volume}{C56}}, \bibinfo{pages}{333} (\bibinfo{year}{2008}),
  \eprint{0804.2741}.

\bibitem[{\citenamefont{Bernabei et~al.}(2010)}]{dama_2010}
\bibinfo{author}{\bibfnamefont{R.}~\bibnamefont{Bernabei}} \bibnamefont{et~al.}
  (\bibinfo{collaboration}{DAMA, LIBRA}), \bibinfo{journal}{Eur. Phys. J.}
  \textbf{\bibinfo{volume}{C67}}, \bibinfo{pages}{39} (\bibinfo{year}{2010}),
  \eprint{1002.1028}.

\bibitem[{\citenamefont{Bernabei et~al.}(2013)}]{dama_2013}
\bibinfo{author}{\bibfnamefont{R.}~\bibnamefont{Bernabei}}
  \bibnamefont{et~al.}, \bibinfo{journal}{Eur. Phys. J.}
  \textbf{\bibinfo{volume}{C73}}, \bibinfo{pages}{2648} (\bibinfo{year}{2013}),
  \eprint{1308.5109}.

\bibitem[{\citenamefont{Scopel and Yoon}(2016)}]{psidm_2015}
\bibinfo{author}{\bibfnamefont{S.}~\bibnamefont{Scopel}} \bibnamefont{and}
  \bibinfo{author}{\bibfnamefont{K.-H.} \bibnamefont{Yoon}},
  \bibinfo{journal}{JCAP} \textbf{\bibinfo{volume}{1602}}, \bibinfo{pages}{050}
  (\bibinfo{year}{2016}), \eprint{1512.00593}.

\bibitem[{\citenamefont{Scopel and Yu}(2017)}]{psidm_2017}
\bibinfo{author}{\bibfnamefont{S.}~\bibnamefont{Scopel}} \bibnamefont{and}
  \bibinfo{author}{\bibfnamefont{H.}~\bibnamefont{Yu}}, \bibinfo{journal}{JCAP}
  \textbf{\bibinfo{volume}{1704}}, \bibinfo{pages}{031} (\bibinfo{year}{2017}),
  \eprint{1701.02215}.

\bibitem[{\citenamefont{Bernabei et~al.}(2018)}]{dama_2018}
\bibinfo{author}{\bibfnamefont{R.}~\bibnamefont{Bernabei}} \bibnamefont{et~al.}
  (\bibinfo{year}{2018}), \eprint{1805.10486}.

\bibitem[{\citenamefont{Kahlhoefer et~al.}(2018)\citenamefont{Kahlhoefer,
  Reindl, Schäffner, Schmidt-Hoberg, and Wild}}]{dama_phase2_first_analysis}
\bibinfo{author}{\bibfnamefont{F.}~\bibnamefont{Kahlhoefer}},
  \bibinfo{author}{\bibfnamefont{F.}~\bibnamefont{Reindl}},
  \bibinfo{author}{\bibfnamefont{K.}~\bibnamefont{Schäffner}},
  \bibinfo{author}{\bibfnamefont{K.}~\bibnamefont{Schmidt-Hoberg}},
  \bibnamefont{and} \bibinfo{author}{\bibfnamefont{S.}~\bibnamefont{Wild}},
  \bibinfo{journal}{JCAP} \textbf{\bibinfo{volume}{1805}}, \bibinfo{pages}{074}
  (\bibinfo{year}{2018}), \eprint{1802.10175}.

\bibitem[{\citenamefont{Baum et~al.}(2018)\citenamefont{Baum, Freese, and
  Kelso}}]{freese_2018}
\bibinfo{author}{\bibfnamefont{S.}~\bibnamefont{Baum}},
  \bibinfo{author}{\bibfnamefont{K.}~\bibnamefont{Freese}}, \bibnamefont{and}
  \bibinfo{author}{\bibfnamefont{C.}~\bibnamefont{Kelso}}
  (\bibinfo{year}{2018}), \eprint{1804.01231}.

\bibitem[{\citenamefont{Kang et~al.}(2018{\natexlab{a}})\citenamefont{Kang,
  Scopel, Tomar, and Yoon}}]{dama_2018_sogang}
\bibinfo{author}{\bibfnamefont{S.}~\bibnamefont{Kang}},
  \bibinfo{author}{\bibfnamefont{S.}~\bibnamefont{Scopel}},
  \bibinfo{author}{\bibfnamefont{G.}~\bibnamefont{Tomar}}, \bibnamefont{and}
  \bibinfo{author}{\bibfnamefont{J.-H.} \bibnamefont{Yoon}},
  \bibinfo{journal}{JCAP} \textbf{\bibinfo{volume}{1807}}, \bibinfo{pages}{016}
  (\bibinfo{year}{2018}{\natexlab{a}}), \eprint{1804.07528}.

\bibitem[{\citenamefont{Herrero-Garcia
  et~al.}(2018)\citenamefont{Herrero-Garcia, Scaffidi, White, and
  Williams}}]{two_comp_dm}
\bibinfo{author}{\bibfnamefont{J.}~\bibnamefont{Herrero-Garcia}},
  \bibinfo{author}{\bibfnamefont{A.}~\bibnamefont{Scaffidi}},
  \bibinfo{author}{\bibfnamefont{M.}~\bibnamefont{White}}, \bibnamefont{and}
  \bibinfo{author}{\bibfnamefont{A.~G.} \bibnamefont{Williams}}
  (\bibinfo{year}{2018}), \eprint{1804.08437}.

\bibitem[{\citenamefont{Adhikari et~al.}(2018{\natexlab{a}})}]{cosine_nature}
\bibinfo{author}{\bibfnamefont{G.}~\bibnamefont{Adhikari}}
  \bibnamefont{et~al.}, \bibinfo{journal}{Nature}
  \textbf{\bibinfo{volume}{564}}, \bibinfo{pages}{83}
  (\bibinfo{year}{2018}{\natexlab{a}}).

\bibitem[{\citenamefont{Aprile et~al.}(2018)}]{xenon_2018}
\bibinfo{author}{\bibfnamefont{E.}~\bibnamefont{Aprile}} \bibnamefont{et~al.}
  (\bibinfo{collaboration}{XENON}) (\bibinfo{year}{2018}), \eprint{1805.12562}.

\bibitem[{\citenamefont{Cui et~al.}(2017)}]{panda_2017}
\bibinfo{author}{\bibfnamefont{X.}~\bibnamefont{Cui}} \bibnamefont{et~al.}
  (\bibinfo{collaboration}{PandaX-II}), \bibinfo{journal}{Phys. Rev. Lett.}
  \textbf{\bibinfo{volume}{119}}, \bibinfo{pages}{181302}
  (\bibinfo{year}{2017}), \eprint{1708.06917}.

\bibitem[{\citenamefont{Akerib et~al.}(2016)}]{lux_complete}
\bibinfo{author}{\bibfnamefont{D.~S.} \bibnamefont{Akerib}}
  \bibnamefont{et~al.} (\bibinfo{year}{2016}), \eprint{1608.07648}.

\bibitem[{\citenamefont{Ahmed et~al.}(2011)}]{cdms_ge}
\bibinfo{author}{\bibfnamefont{Z.}~\bibnamefont{Ahmed}} \bibnamefont{et~al.}
  (\bibinfo{collaboration}{CDMS-II}), \bibinfo{journal}{Phys. Rev. Lett.}
  \textbf{\bibinfo{volume}{106}}, \bibinfo{pages}{131302}
  (\bibinfo{year}{2011}), \eprint{1011.2482}.

\bibitem[{\citenamefont{Agnese et~al.}(2014{\natexlab{a}})}]{cdms_lite}
\bibinfo{author}{\bibfnamefont{R.}~\bibnamefont{Agnese}} \bibnamefont{et~al.}
  (\bibinfo{collaboration}{SuperCDMS}), \bibinfo{journal}{Phys. Rev. Lett.}
  \textbf{\bibinfo{volume}{112}}, \bibinfo{pages}{041302}
  (\bibinfo{year}{2014}{\natexlab{a}}), \eprint{1309.3259}.

\bibitem[{\citenamefont{Agnese et~al.}(2014{\natexlab{b}})}]{super_cdms}
\bibinfo{author}{\bibfnamefont{R.}~\bibnamefont{Agnese}} \bibnamefont{et~al.}
  (\bibinfo{collaboration}{SuperCDMS}), \bibinfo{journal}{Phys. Rev. Lett.}
  \textbf{\bibinfo{volume}{112}}, \bibinfo{pages}{241302}
  (\bibinfo{year}{2014}{\natexlab{b}}), \eprint{1402.7137}.

\bibitem[{\citenamefont{Agnese et~al.}(2015)}]{cdms_2015}
\bibinfo{author}{\bibfnamefont{R.}~\bibnamefont{Agnese}} \bibnamefont{et~al.}
  (\bibinfo{collaboration}{SuperCDMS}), \bibinfo{journal}{Phys. Rev.}
  \textbf{\bibinfo{volume}{D92}}, \bibinfo{pages}{072003}
  (\bibinfo{year}{2015}), \eprint{1504.05871}.

\bibitem[{\citenamefont{Ullio et~al.}(2001)\citenamefont{Ullio, Kamionkowski,
  and Vogel}}]{spin_n_suppression}
\bibinfo{author}{\bibfnamefont{P.}~\bibnamefont{Ullio}},
  \bibinfo{author}{\bibfnamefont{M.}~\bibnamefont{Kamionkowski}},
  \bibnamefont{and} \bibinfo{author}{\bibfnamefont{P.}~\bibnamefont{Vogel}},
  \bibinfo{journal}{JHEP} \textbf{\bibinfo{volume}{07}}, \bibinfo{pages}{044}
  (\bibinfo{year}{2001}), \eprint{hep-ph/0010036}.

\bibitem[{\citenamefont{Del~Nobile et~al.}(2015)\citenamefont{Del~Nobile,
  Gelmini, Georgescu, and Huh}}]{spin_gelmini}
\bibinfo{author}{\bibfnamefont{E.}~\bibnamefont{Del~Nobile}},
  \bibinfo{author}{\bibfnamefont{G.~B.} \bibnamefont{Gelmini}},
  \bibinfo{author}{\bibfnamefont{A.}~\bibnamefont{Georgescu}},
  \bibnamefont{and} \bibinfo{author}{\bibfnamefont{J.-H.} \bibnamefont{Huh}},
  \bibinfo{journal}{JCAP} \textbf{\bibinfo{volume}{1508}}, \bibinfo{pages}{046}
  (\bibinfo{year}{2015}), \eprint{1502.07682}.

\bibitem[{\citenamefont{Anand et~al.}(2014)\citenamefont{Anand, Fitzpatrick,
  and Haxton}}]{haxton2}
\bibinfo{author}{\bibfnamefont{N.}~\bibnamefont{Anand}},
  \bibinfo{author}{\bibfnamefont{A.~L.} \bibnamefont{Fitzpatrick}},
  \bibnamefont{and} \bibinfo{author}{\bibfnamefont{W.~C.}
  \bibnamefont{Haxton}}, \bibinfo{journal}{Phys. Rev.}
  \textbf{\bibinfo{volume}{C89}}, \bibinfo{pages}{065501}
  (\bibinfo{year}{2014}), \eprint{1308.6288}.

\bibitem[{\citenamefont{Behnke et~al.}(2012)}]{coupp}
\bibinfo{author}{\bibfnamefont{E.}~\bibnamefont{Behnke}} \bibnamefont{et~al.}
  (\bibinfo{collaboration}{COUPP}), \bibinfo{journal}{Phys. Rev.}
  \textbf{\bibinfo{volume}{D86}}, \bibinfo{pages}{052001}
  (\bibinfo{year}{2012}), \bibinfo{note}{[Erratum: Phys.
  Rev.D90,no.7,079902(2014)]}, \eprint{1204.3094}.

\bibitem[{\citenamefont{Behnke et~al.}(2017)}]{picasso}
\bibinfo{author}{\bibfnamefont{E.}~\bibnamefont{Behnke}} \bibnamefont{et~al.},
  \bibinfo{journal}{Astropart. Phys.} \textbf{\bibinfo{volume}{90}},
  \bibinfo{pages}{85} (\bibinfo{year}{2017}), \eprint{1611.01499}.

\bibitem[{\citenamefont{Amole et~al.}(2017)}]{pico60}
\bibinfo{author}{\bibfnamefont{C.}~\bibnamefont{Amole}} \bibnamefont{et~al.}
  (\bibinfo{collaboration}{PICO}), \bibinfo{journal}{Phys. Rev. Lett.}
  \textbf{\bibinfo{volume}{118}}, \bibinfo{pages}{251301}
  (\bibinfo{year}{2017}), \eprint{1702.07666}.

\bibitem[{\citenamefont{Amole et~al.}(2015)}]{pico2l}
\bibinfo{author}{\bibfnamefont{C.}~\bibnamefont{Amole}} \bibnamefont{et~al.}
  (\bibinfo{collaboration}{PICO}), \bibinfo{journal}{Phys. Rev. Lett.}
  \textbf{\bibinfo{volume}{114}}, \bibinfo{pages}{231302}
  (\bibinfo{year}{2015}), \eprint{1503.00008}.

\bibitem[{\citenamefont{Tucker-Smith and Weiner}(2001)}]{inelastic}
\bibinfo{author}{\bibfnamefont{D.}~\bibnamefont{Tucker-Smith}}
  \bibnamefont{and} \bibinfo{author}{\bibfnamefont{N.}~\bibnamefont{Weiner}},
  \bibinfo{journal}{Phys. Rev.} \textbf{\bibinfo{volume}{D64}},
  \bibinfo{pages}{043502} (\bibinfo{year}{2001}), \eprint{hep-ph/0101138}.

\bibitem[{\citenamefont{Kang et~al.}(2018{\natexlab{b}})\citenamefont{Kang,
  Scopel, Tomar, and Yoon}}]{sensitivities_2018}
\bibinfo{author}{\bibfnamefont{S.}~\bibnamefont{Kang}},
  \bibinfo{author}{\bibfnamefont{S.}~\bibnamefont{Scopel}},
  \bibinfo{author}{\bibfnamefont{G.}~\bibnamefont{Tomar}}, \bibnamefont{and}
  \bibinfo{author}{\bibfnamefont{J.-H.} \bibnamefont{Yoon}}
  (\bibinfo{year}{2018}{\natexlab{b}}), \eprint{1805.06113}.

\bibitem[{\citenamefont{Koposov et~al.}(2010)\citenamefont{Koposov, Rix, and
  Hogg}}]{v0_koposov}
\bibinfo{author}{\bibfnamefont{S.~E.} \bibnamefont{Koposov}},
  \bibinfo{author}{\bibfnamefont{H.-W.} \bibnamefont{Rix}}, \bibnamefont{and}
  \bibinfo{author}{\bibfnamefont{D.~W.} \bibnamefont{Hogg}},
  \bibinfo{journal}{Astrophys. J.} \textbf{\bibinfo{volume}{712}},
  \bibinfo{pages}{260} (\bibinfo{year}{2010}), \eprint{0907.1085}.

\bibitem[{\citenamefont{Piffl et~al.}(2014)}]{vesc_2014}
\bibinfo{author}{\bibfnamefont{T.}~\bibnamefont{Piffl}} \bibnamefont{et~al.},
  \bibinfo{journal}{Astron. Astrophys.} \textbf{\bibinfo{volume}{562}},
  \bibinfo{pages}{A91} (\bibinfo{year}{2014}), \eprint{1309.4293}.

\bibitem[{\citenamefont{Bernabei et~al.}(1998)}]{dama_1998}
\bibinfo{author}{\bibfnamefont{R.}~\bibnamefont{Bernabei}}
  \bibnamefont{et~al.}, \bibinfo{journal}{Phys. Lett.}
  \textbf{\bibinfo{volume}{B424}}, \bibinfo{pages}{195} (\bibinfo{year}{1998}).

\bibitem[{\citenamefont{Fox et~al.}(2011)\citenamefont{Fox, Liu, and
  Weiner}}]{factorization}
\bibinfo{author}{\bibfnamefont{P.~J.} \bibnamefont{Fox}},
  \bibinfo{author}{\bibfnamefont{J.}~\bibnamefont{Liu}}, \bibnamefont{and}
  \bibinfo{author}{\bibfnamefont{N.}~\bibnamefont{Weiner}},
  \bibinfo{journal}{Phys. Rev.} \textbf{\bibinfo{volume}{D83}},
  \bibinfo{pages}{103514} (\bibinfo{year}{2011}), \eprint{1011.1915}.

\bibitem[{\citenamefont{Del~Nobile et~al.}(2013)\citenamefont{Del~Nobile,
  Gelmini, Gondolo, and Huh}}]{del_nobile_generalized}
\bibinfo{author}{\bibfnamefont{E.}~\bibnamefont{Del~Nobile}},
  \bibinfo{author}{\bibfnamefont{G.}~\bibnamefont{Gelmini}},
  \bibinfo{author}{\bibfnamefont{P.}~\bibnamefont{Gondolo}}, \bibnamefont{and}
  \bibinfo{author}{\bibfnamefont{J.-H.} \bibnamefont{Huh}},
  \bibinfo{journal}{JCAP} \textbf{\bibinfo{volume}{1310}}, \bibinfo{pages}{048}
  (\bibinfo{year}{2013}), \eprint{1306.5273}.

\bibitem[{\citenamefont{Kang et~al.}(2018{\natexlab{c}})\citenamefont{Kang,
  Scopel, Tomar, Yoon, and Gondolo}}]{anapole_2018}
\bibinfo{author}{\bibfnamefont{S.}~\bibnamefont{Kang}},
  \bibinfo{author}{\bibfnamefont{S.}~\bibnamefont{Scopel}},
  \bibinfo{author}{\bibfnamefont{G.}~\bibnamefont{Tomar}},
  \bibinfo{author}{\bibfnamefont{J.-H.} \bibnamefont{Yoon}}, \bibnamefont{and}
  \bibinfo{author}{\bibfnamefont{P.}~\bibnamefont{Gondolo}}
  (\bibinfo{year}{2018}{\natexlab{c}}), \eprint{1808.04112}.

\bibitem[{\citenamefont{Adhikari et~al.}(2018{\natexlab{b}})}]{cosine_bck}
\bibinfo{author}{\bibfnamefont{P.}~\bibnamefont{Adhikari}} \bibnamefont{et~al.}
  (\bibinfo{collaboration}{COSINE-100}), \bibinfo{journal}{Eur. Phys. J.}
  \textbf{\bibinfo{volume}{C78}}, \bibinfo{pages}{490}
  (\bibinfo{year}{2018}{\natexlab{b}}), \eprint{1804.05167}.

\bibitem[{cos()}]{cosine_private}
\bibinfo{howpublished}{COSINE--100 Collaboration, private communication}.

\end{thebibliography}
\end{document}